\documentclass{aa}  

\usepackage{graphicx}
\usepackage{txfonts}
\usepackage{hyperref}
\hypersetup{
    final=true,
    pageanchor=true,
    colorlinks=true,
    breaklinks=true,
    linkcolor=blue,
    citecolor=blue,
    urlcolor=blue,
    pdfpagemode=UseNone}

\usepackage[switch, modulo]{lineno}
\usepackage{natbib}
\bibpunct{(}{)}{;}{a}{}{,}
\usepackage{multirow}
\usepackage{nicefrac}
\usepackage{color}

\definecolor{Red}{rgb}{.9,.1,.1}

\newcommand\pha{\phantom{0}}
\newcommand\phb{\phantom{00}}
\newcommand\phc{\phantom{$-$.}}
\newcommand\phd{\phantom{$-$}}
\newcommand\phe{\phantom{$-$0.}}

\begin{document}

   \title{Ca \textsc{ii} 8542\,\AA\ brightenings induced by a solar microflare}

   \author{C. Kuckein\inst{1}
          \and
          A. Diercke\inst{1,2}
          \and
          S. J. Gonz\'alez Manrique\inst{3,1,2}
          \and  
          M. Verma\inst{1}
          \and \\
          J. L\"ohner-B\"ottcher\inst{4}
          \and
          H. Socas-Navarro\inst{5}
          \and
          H. Balthasar\inst{1}
          \and
          M. Sobotka\inst{6}
          \and 
          C. Denker\inst{1}
          }

   \institute{Leibniz-Institut f\"ur Astrophysik Potsdam (AIP),
              An der Sternwarte 16,14482 Potsdam, Germany\\
              \email{ckuckein@aip.de}
          \and
          Universit{\"a}t Potsdam, Institut f{\"u}r Physik und Astronomie, Karl-Liebknecht-Str. 24/25, 14476 Potsdam, Germany
          \and
          Astronomical Institute of the Slovak Academy of Sciences, 05960, Tatransk\'a Lomnica, Slovak Republic
          \and
          Kiepenheuer-Institut f{\"u}r Sonnenphysik, Sch{\"o}neckstr. 6, 79104 Freiburg, Germany
          \and
          Instituto de Astrof\'isica de Canarias, Avda v\'ia L\'actea S/N, 38205 La Laguna, Tenerife, Spain          
          \and
          Astronomical Institute, Academy of Sciences of the Czech Republic, Fri\v{c}ova 298, 
          25165 Ond\v{r}ejov, Czech Republic  
           }

   \date{Version: \today }

% \abstract{}{}{}{}{} 
% 5 {} token are mandatory
 
  \abstract
   {} %leave it empty if necessary  
   {We study small-scale brightenings  in \ion{Ca}{ii} 8542\,\AA\ line-core images to determine their
   nature and effect on localized heating and mass transfer in active regions.}
   {High-resolution two-dimensional spectroscopic observations of a solar active region in the near-infrared 
   \ion{Ca}{ii} 8542\,\AA\ line were acquired with the GREGOR Fabry-P\'erot Interferometer
   attached to the 1.5-meter GREGOR telescope. 
   Inversions of the spectra were 
   carried out using the NICOLE code to infer temperatures and line-of-sight (LOS) velocities. 
   Response functions of the \ion{Ca}{ii} line were computed for temperature and LOS velocity variations.
   Filtergrams of the Atmospheric Imaging Assembly (AIA) and magnetograms of the Helioseismic and Magnetic Imager (HMI)
   were coaligned to match the ground-based observations and to follow the 
   \ion{Ca}{ii} brightenings along all available layers of the atmosphere.
     }
   {We identified three brightenings of sizes up to 2\arcsec\,$\times$\,2\arcsec\ that appeared in 
   the \ion{Ca}{ii} 8542\,\AA\ line-core images. Their lifetimes were at least 1.5\,min.
   We found evidence that the brightenings belonged to the footpoints of a microflare (MF).
   The properties of the observed brightenings disqualified the scenarios of Ellerman bombs or 
   Interface Region Imaging Spectrograph (IRIS) bombs. 
   However, this MF shared some common properties with flaring active-region fibrils or flaring arch filaments (FAFs):
   (1) FAFs and MFs are both apparent in chromospheric and coronal layers according to the AIA channels, 
   and (2) both show flaring arches with lifetimes of about 3.0--3.5\,min and lengths of $\sim$\,20\arcsec\
   next to the brightenings. 
   The inversions revealed heating by 600\,K at the footpoint location in the ambient chromosphere during the impulsive 
   phase.
   Connecting the footpoints, a dark filamentary structure appeared in 
   the \ion{Ca}{ii} line-core images. 
   Before the start of the MF, the spectra of this structure already indicated average blueshifts, 
   meaning upward motions of the plasma along the LOS. 
   During the impulsive phase, these velocities increased up to $-2.2$\,km\,s$^{-1}$. The structure did not disappear
   during the observations. Downflows dominated at the footpoints.
   However, in the upper photosphere, slight upflows occurred during the impulsive phase. 
   Hence, bidirectional flows are present in the footpoints of the MF.
   }
   {We detected \ion{Ca}{ii} brightenings that coincided with the footpoint location of an MF. The MF event 
   led to a rise of plasma in the upper photosphere, both before and during the impulsive phase. Excess  mass, previously raised to at most chromospheric layers,  slowly drained downward along arches toward the footpoints of the MF.
  }


   \authorrunning{Kuckein et al.}
   \titlerunning{Ca \textsc{ii} 8542\,\AA\ brightenings induced by a solar microflare}      
             
   \maketitle

% =====================================================================
\section{Introduction}
% =====================================================================
Small-scale and intense short-lived brightenings are often seen in the 
solar atmosphere. 
These impulsive events are frequently 
hypothesized to form as a result of magnetic reconnection, 
which if true would mean that they also produce local heating.
In the past, such events were mainly studied with ground-based facilities.
Today, space-borne telescopes have become powerful tools 
especially in combination
with new high-resolution ground-based data. This ensures unprecedented 
multiwavelength analyses of impulsive events across many layers of the 
solar atmosphere. 

There are several types of similar-looking brightenings, which increases the difficulty 
of understanding these phenomena and their relation 
\citep[e.g.,][]{rutten13}. 
Fundamental differences in their 
lifetime, size, location, expansion along the atmosphere, and other characteristic observational signatures
distinguish these events, however.

The by far most frequently studied
small-scale brightenings are Ellerman bombs (EBs), which have
been introduced by \citet[][]{ellerman17}
as `solar hydrogen bombs'. From an observational point of view, they are intensity enhancements
of the H$\alpha$ line wings, while the line core remains unaffected. The same applies to the
\ion{Ca}{ii} 8542\,\AA\ line \citep{fang06b}, even though exceptions are possible \citep{rezaei15}. 
No signature is observed higher up in the corona \citep{schmieder04}.
Together with evidence that EBs are best identified
in 1600 and 1700\,\AA\ filtergrams \citep{berlicki10,pariat07,vissers13}, these facts suggest 
that they are located in the photosphere \citep{rutten13}, although discrepancies about their formation height 
still exist \citep[e.g.,][]{berlicki14}. 
Their lifetimes are usually shorter
than 30\,min, predominantly between 1.5 and 7\,min with a peak between 3 and 4\,min \citep{pariat07, vissers13}, 
although longer-lived EBs have been detected \citep{bello13}.
EBs are typically located in active regions, surrounded by magnetic field
concentrations or at polarity inversion lines
\citep[e.g.,][]{matsumoto08,nelson13,reid16}, and
have a small elongated shape with typical sizes of $1\farcs8 \times 1\farcs1$ \citep{georgoulis02}. 
EBs lead to temperature enhancements of up to 600--1300\,K, derived from observational \ion{Ca}{ii} 8542\,\AA\ 
data \citep{fang06b}. Recent ground-based high-resolution observations probably indicate that even 
smaller EBs exist (0\farcs3--0\farcs8) with temperature increments of up to 3000\,K \citep{li15}.
Owing to these previous results, it is widely considered that EBs result from magnetic reconnection \citep[][and references therein]{vissers13}.

An extensive amount of literature has been published on microflares (MFs). They are small highly energetic brightenings. 
MFs are often observed as hard X-ray bursts,
are short lived \citep[2.2--15\,min;][]{hannah08}, 
and can occur every 5\,min in active regions during solar maximum \citep{lin84}. 
Moreover, precursor brightenings have been detected in light curves of 1600\,\AA\ \citep{brosius09}.
The apparent chromospheric counterpart of an MF in H$\alpha$ \citep{canfield87,liu04} shows fine structure and 
sizes of a few arcseconds \citep{berkebile09}. 
\citet{fang06a} reported on emission cores in H$\alpha$ and \ion{Ca}{ii} 8542\,\AA\ intensity spectra 
during MFs. These authors inferred temperature increments of about 1000--2200\,K in the lower chromosphere during the MFs. 
This increment is in accordance with the observed line-core emission of \ion{Ca}{ii} 8542\,\AA. 
High bidirectional velocities between $\pm$\,(40--50)\,km\,s$^{-1}$ were derived using the \ion{Ca}{ii} 8542\,\AA\
line \citep{hong16}. Microflares are frequently found near magnetic polarity inversion lines \citep{liu04}.

A new class of brightening was recently discovered with the 
space-borne Interface Region Imaging Spectrograph \citep[IRIS;][]{depontieu14}. 
The so-called IRIS bombs (IBs) consist of small roundish pockets of photospheric hot 
gas. \citet{peter14} reported on transient features with
lifetimes of about 5\,min that occur in active regions where magnetic fields cancel. 
These brightenings resemble EBs, with the difference that they manifest in much higher photospheric temperatures 
(up to $100\,000$\,K). 
In an independent analysis, \citet{judge15} proposed that these bombs are rather located higher, in the 
low- to mid-chromosphere. Nevertheless, it has been shown that some EBs are co-spatial to IBs 
\citep{tian16}. 
As a consequence, if IBs are directly related to EBs, existing models might underestimate the temperature 
increments of the latter \citep[<3000\,K,][]{li15}. 
The formation height of EBs might play an important role when
determining their temperature. In coronal filtergrams from the Atmospheric Imaging Assembly \citep[AIA,][]{aia}
no indication of IBs are found.

\begin{figure}[!t]
 \centering
 \includegraphics[width=\hsize]{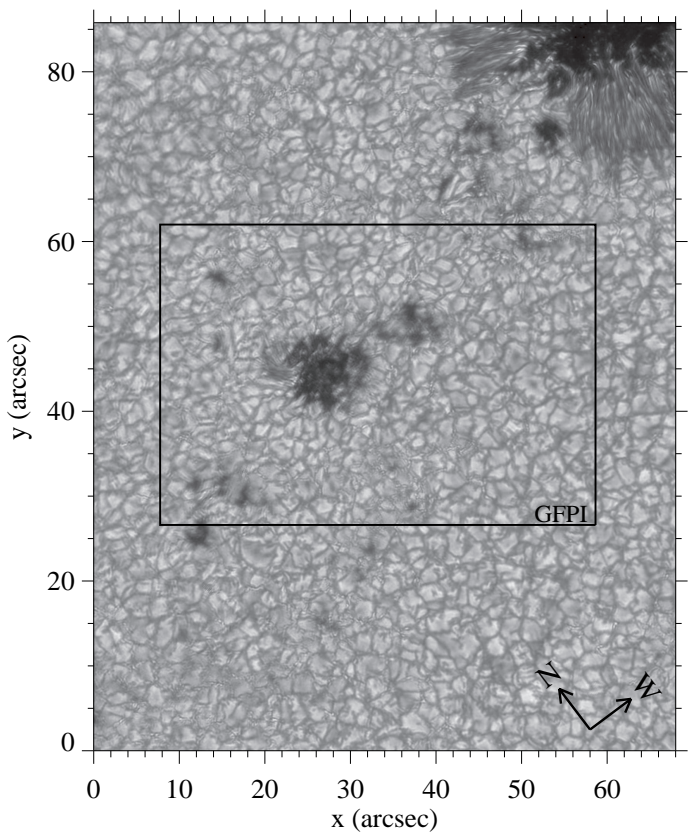}
\caption{Speckle-restored blue continuum (4505\,\AA) image of active region NOAA 12149 at 08:38\,UT 
 on 2014 August 26 observed with the PCO.4000 CCD camera installed in the 
 blue imaging channel of the GFPI at GREGOR. 
 The rectangle outlines the FOV of 
 the  GFPI. Arrows in the bottom right-hand corner
 indicate solar north and west.} 
  \label{Fig:BIC_overview}
\end{figure}

Few data are published on flaring arch filaments. \citet{vissers15} defined them 
as small brightenings with an elongated morphology that
have shorter lifetimes than EBs. However, \citet{rutten16} suggested a more accurate name for these events: "flaring 
active-region fibrils" (FAFs), in order to avoid confusion with larger filaments. 
FAFs are well identified in 1600\,\AA\ filtergrams, while their appearance, in contrast 
to EBs, is not always clear in 1700\,\AA\ images. \citet{vissers15} studied two FAFs in detail and concluded that 
(1) FAFs occur in active regions, 
(2) they probably indicate magnetic reconnection but higher 
in the atmosphere than EBs, 
(3) they show a fibril-like morphology, and 
(4) they produce bright arches in the hotter chromosphere and corona. 
\citet{pariat09} reported ``transient loops'' in 1600\,\AA\ images 
that probably are FAFs. In the past, FAFs were already studied, but no common name was suggested so far.
As pointed out by \citet{rutten16}, FAFs are very similar to EBs with the difference that they also affect the 
higher atmospheric layers. 

In this work, we report on high-resolution ground-based observations using \ion{Ca}{ii} 8542\,\AA\ imaging spectra
in an active region. The observations include three small brightening areas of the near-infrared
\ion{Ca}{ii} line core that are carefully discussed and placed into context using space observations.
Temperature and Doppler velocity statistics inferred from spectral-line inversions are presented.

\begin{figure*}[!t] 
   \resizebox{\hsize}{!}{\includegraphics{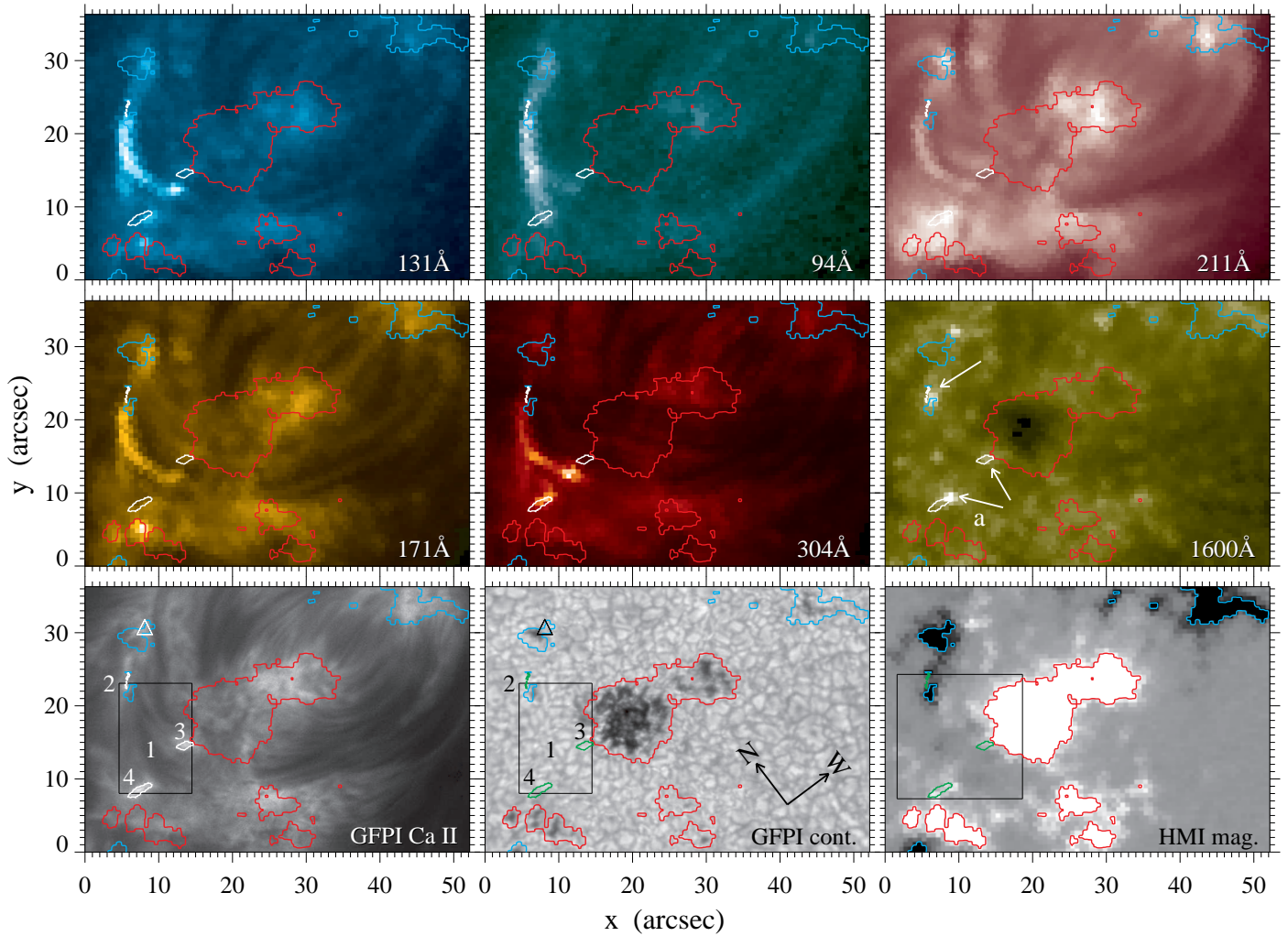}}
      \caption{Active region NOAA 12149 observed with SDO and GFPI. SDO images are 
      shown for the time period 08:37:23 -- 08:37:50\,UT. The GFPI filtergrams correspond to 08:38:13\,UT. 
      Blue and red contours mark the negative and positive polarities, respectively, obtained from the HMI 
      magnetogram saturated to $\pm 350$\,G in the lower right panel. 
      The white contours (green in the case of the two bottom rightmost images), 
      which are enumerated with numbers 2--4,
      correspond to the brightenings 
      in the GFPI \ion{Ca}{II} line-core image between 08:37--08:39\,UT. 
      Number 1 marks an area of dark filamentary structure as seen in the \ion{Ca}{II} line-core image.
      Solar north and west are
          indicated by the arrows in the GFPI continuum panel at the bottom. The box in the HMI panel shows the
          FOV of the magnetograms in Fig. \ref{Fig:magnetograms}. The rectangle in the GFPI panels outlines the 
          FOV of the inversions shown in Figs. \ref{Fig:temperatures} and \ref{Fig:velocities}. 
          The triangle in the upper left corner of the GFPI panels highlights a persistent enhancement of 
          the \ion{Ca}{ii} line core.
          The arrows in the AIA 1600\,\AA\ panel indicate three areas with sudden brightenings.
          The temporal evolution of this figure is shown between 08:35:08 and 08:39:08 in an online movie.}
\label{Fig:overview}
\end{figure*}

% =====================================================================
\section{Observations}
% =====================================================================

The observations were carried out on 2014 August 26 with the 1.5-meter GREGOR solar telescope \citep{schmidt12} 
located on Tenerife, Spain. 
We observed a group of pores in active region NOAA 12149 at coordinates ($x,y$)\,$=$\,($-240\arcsec,40\arcsec$)
and cosine of heliocentric angle $\mu$\,$\sim$\,0.97, close to the solar disk center (Fig. \ref{Fig:BIC_overview}).  
The pores show opposite polarity to the leading sunspot.
The primary instrument was the GREGOR Fabry-P\'erot Interferometer (GFPI).
The GFPI is equipped with two synchronized cameras. 
One records continuum images in the broad-band channel, 
the other observes spectral narrow-band images filtered by two etalons in the collimated beam.
We refer to \citet[][]{puschmann12} and references therein for
a complete overview of the GFPI.
The setup allows for image restoration using multi-object multi-frame blind deconvolution
\citep[MOMFBD,][]{lofdahl02,vannoort05}.

The GFPI was operated in the imaging spectroscopic mode with a prefilter centered at the 
\ion{Ca}{ii} 8542\,\AA\ line. Since the quantum
efficiency of the detector substantially decreases when approaching the upper limit of 8600\,\AA, 
exposure times of 80\,ms were necessary. Four accumulations were 
performed for each of the selected 36 wavelength positions.
A nonuniform step size was chosen to scan the \ion{Ca}{ii} line.
The wings were sampled wider with a step size of 200\,m\AA,\ while the inner wings and line core were
sampled narrower with 80\,m\AA, spanning a total spectral window of 4.6\,\AA.
The individual spectral positions with respect to the \ion{Ca}{ii} 8542\,\AA\ line-core center 
were $-$2.22, $-$2.02, $-$1.82, $-$1.62, $-$1.41, $-$1.21, 
   $-$1.01, $-$0.81, $-$0.73, $-$0.65, $-$0.56, $-$0.48, $-$0.40, $-$0.32, $-$0.24, $-$0.16, $-$0.08, 
   0.00, 0.08, 0.16, 0.24, 0.32, 0.40, 0.48, 0.56, 0.65, 0.73, 0.81, 1.01, 1.21, 1.41, 1.62, 
   1.82, 2.02, 2.22, and 2.42~\AA.

We acquired 20 scan sequences, starting at 08:23:57\,UT and ending
at 08:39:25\,UT, 
with a cadence of 47\,s. The field of view (FOV) of the GFPI was $52\arcsec \times 36\arcsec$, with an 
image scale of $\sim$\,0\farcs081\,pixel$^{-1}$ in the 2$\times$2-pixel binning mode.

A context image is provided by the PCO.4000 CCD camera installed in the blue imaging channel of the GFPI (Fig. \ref{Fig:BIC_overview}).  
The camera was equipped with a blue continuum filter (4505\,\AA) and acquired
80 images in one burst. 
The FOV was $68\arcsec \times 86\arcsec$, with an image scale of about 0\farcs035\,pixel$^{-1}$. 

In support of our ground-based observations, we analyzed data from the Solar Dynamics Observatory \citep[SDO,][]{sdo}.
We focused on filtergrams recorded by AIA in the 
wavelength bands 131, 
94, 335, 211, 193, 171, 304, 1600, and 1700\,\AA. These wavelengths
give us the opportunity to track any event across different temperatures and heights in the solar atmosphere, 
from the photosphere to the corona. The cadence for
1600 and 1700\,\AA\ data was 24\,s, while all other channels were acquired with a cadence of 12\,s.  Furthermore, 
magnetograms from the Helioseismic and Magnetic Imager \citep[HMI,][]{hmi} with a cadence of 45\,s were used to 
complement the filtergrams with information of the line-of-sight (LOS) magnetic field.

% =====================================================================
\section{Data reduction and analysis}
% =====================================================================
The ground-based data were reduced using the reduction pipeline `sTools' 
\citep[][]{stools}. It included dark and flat-field corrections for both instruments. In addition, 
for GFPI, blueshift, prefilter transmission
corrections, and a continuum adjustment of the spectral profile by matching it with an atlas profile from the 
Fourier transform spectra \citep[FTS,][]{neckel84} spectrometer were performed. 
The image restoration of GFPI's broad-band channel using MOMFBD was successful. Unfortunately, the algorithm delivered unsatisfactory \ion{Ca}{ii} 8542\,\AA\ narrow-band images. We ascribe this to the relatively low signal-to-noise ratio of the data. Nevertheless, a comparison between the spectral profiles arising from the restored and non-restored images in our data set showed negligible changes. 
Therefore our study was carried out with the non-restored \ion{Ca}{ii} spectroscopic images.
The speckle-interferometry code KISIP \citep{woeger08} was used to restore the blue continuum images.
Image rotation induced by the altitude-azimuth mount of the telescope \citep{volkmer12} was removed.

%--------------------------------------------------------------------
\subsection{GFPI wavelength calibration}
We calculated the median intensity profile 
inside an area of 135\,$\times$\,240 pixels ($\sim$\,11\arcsec\,$\times$\,19\arcsec). 
This area was located in the right-hand bottom part of the FOV, in the absence of 
magnetic structures as seen in the HMI magnetogram.
A polynomial fit to the line core of this profile yielded the central position.
We considered the central position as the wavelength at rest of the 
\ion{Ca}{ii} line, which we established at 8542.09\,\AA\ \citep{nist}. 
The theoretical dispersion of the GFPI was 4.04\,m\AA\,position$^{-1}$
at 8542\,\AA. We computed the wavelength array by
multiplying the 36 scan positions along the line by the dispersion. 
We are limited to this calibration since a more
accurate wavelength calibration needed nearby telluric lines or synthetic laser lines.

%--------------------------------------------------------------------
\subsection{Solar Dynamics Observatory}
We started with processed Level 1.0 data from AIA and HMI for the time period between 08:00--09:00\,UT. Level 1.0 implies 
a basic data reduction, that is, flat-field correction and conversion into FITS format \citep{aia}. 
Further processing steps are carried out with the `Solar Software' routines \citep[SSW,][]{freeland98,bentely98}, 
like the adjustment of the image scale of HMI to AIA (\mbox{0\farcs5} $\rightarrow$ \mbox{0\farcs6\,pixel$^{-1}$}) \citep{sdoguide2013}. 
The alignment of the data was carried out with the SSW routines to derotate the 
images with respect to the central image of the time series at 08:30\,UT. 

All SDO data were then coaligned to the GFPI FOV (Fig.~\ref{Fig:overview}). This is achieved by vertical mirroring, 
magnification, rotation ($\sim$37$^\circ$), and shifting of the corresponding FOV of SDO.

%--------------------------------------------------------------------
\subsection{\ion{Ca}{ii} 8542\,\AA\ inversions }\label{Sect:inversions}
To deeper investigate the \ion{Ca}{ii} spectra, we used the inversion code NICOLE 
\citep[][]{socas15} to infer the physical parameters of the atmosphere where the line was formed. 
This includes temperatures and LOS velocities at several optical depths, that is, at different heights.
The code carries out non-local thermodynamical equilibrium (NLTE) inversions of 
Zeeman-induced Stokes profiles. 
Its robustness and good performance using the \ion{Ca}{ii} 8542\,\AA\ line was 
demonstrated in recent works \citep[e.g.,][]{delacruz12,delacruz13,leenaarts14,quinteronoda16}. 
Since we only recorded intensity profiles (Stokes $I$), we set
the other Stokes parameters ($Q$, $U$, and $V$) to zero as well as their weights.  

NICOLE needs an initial-guess atmosphere to compute the arising intensity profiles and compare them with the
observed ones. 
Several initial atmospheres were tested, and the best 
fits to the observed intensity profiles were
achieved with the Harvard-Smithsonian reference atmosphere \citep[HSRA,][]{gingerich71}. The HSRA covers
a wide range of heights, which are defined in terms of the logarithm of the LOS continuum optical depth $\tau$ 
at a wavelength of 5000\,\AA,\ and it spans the interval between $-8.0$\,$\leq$\,$\log \tau$\,$\leq$\,1.4. 
Hence, it extends 
from the photosphere to the transition region. 
The initial macroturbulence was set to 2\,km\,s$^{-1}$, to ensure smoother synthesized profiles, 
and was left as degree of freedom. 
We weighted the 36 spectral positions to concentrate only on the 23 points in the inner part of the 
\ion{Ca}{ii} line, in the range $-$0.8\,$\leq$\,$\Delta \lambda$\,$\leq$\,1.2\,\AA\ from the line center. 
The outer wings of the spectral line were omitted, only taking the outermost point on each side into account.
This was motivated by the relatively slow cadence of one spectral scan (47\,s). 
In the presence of dynamic events, 
the spectral profile might undergo substantial changes during the scan. 
By concentrating only on the line core and inner wings, we achieved better fits with NICOLE 
and still ensured a reliable temperature stratification. 

Following the recommendation of \citet{leenaarts14}, who reported on the effects of various isotopes
of Ca with slightly different energies, we included the effects of isotopic splitting in NICOLE. 
We used up to 11 nodes for the temperature and 8 nodes for the LOS velocity.
This version of NICOLE is unique compared to other similar inversion codes in that it uses a 
regularization scheme that ensures that smoother solutions are preferred. Thus, we only needed to 
ensure that the number of nodes was sufficiently large to provide the freedom necessary to fit the profiles. 
In practice, however, the number of degrees of freedom is smaller than the number of nodes because the regularization effectively reduces the freedom. Therefore, if the number of nodes is large enough, reducing it will 
not change the solution significantly until the nodes are too few and there is not 
enough freedom to fit the observations.
The magnetic field was not taken into account. 
In addition, 
three inversion sequences with five, three, and again three inversion attempts each, respectively, 
were chosen as part of the inversion strategy. 
The goodness of the fits were evaluated using
the $\chi^2$ value, which is defined as the sum of the squared difference between the observed and the 
synthesized intensity profiles. The lower the $\chi^2$ value, the better the fit. 
The resulting atmosphere from the first inversion was
used as input for the second inversion sequence, and so on.  
As expected, mostly, but not always, the third inversion sequence provided the lowest $\chi^2$ and hence best result. 
Moreover, a visual inspection of the fits to the observed spectra was 
carried out to ensure the code's performance. 
We kept the best result of the three cycles with the lowest $\chi^2$ value. 
A threshold of $\chi^2 \le 0.008$ was chosen after 
visually examining very many random fits to ensure good-quality fits. 
Inversions with larger $\chi^2$ were dropped (less than 3.5\,\% of the pixels). 
In the physical maps in Figs. \ref{Fig:temperatures} and \ref{Fig:velocities}, 
the excluded pixels were substituted by the 
median of the surrounding $3 \times 3$ pixel mosaic.  

An inversion of the whole FOV in all available maps was not possible. 
The inversion strategy for the NLTE inversions with NICOLE was, from a computational point of view, 
very time consuming. However, it provided the most reliable results. 
Therefore, we concentrated on a smaller FOV of $123\times 187$ pixels ($\sim$\,10\arcsec\,$\times$\,15\arcsec), 
which included the most relevant 
features in the \ion{Ca}{ii} line-core and SDO images. 
In this smaller FOV, we monitored the changes in temperature and LOS velocity. The FOV is outlined with a black rectangle
at the GFPI panels in the bottom of Fig. \ref{Fig:overview}.

\subsubsection{Response functions}\label{Sect:responsefunct}
The \ion{Ca}{ii} 8542\,\AA\ spectral line extends across several layers of the solar atmosphere. The line core
lies in the chromosphere, whereas the outer wings reach the photosphere. 
By computing response functions (RFs) 
to perturbations of the temperature and LOS velocity of a given atmospheric model, 
we infer at which atmospheric heights (in $\log\tau$ units) the \ion{Ca}{ii} line changes most noticeably in that model.
The larger the RF, the higher the sensitivity of the spectral line to perturbations of a given 
physical quantity at a specific height. 
This allows focusing our analysis of the temperatures and LOS velocities on the relevant height ranges.

We followed the approach of \citet{quinteronoda16} to
compute numerical RFs by perturbing the atmosphere at each $\log\tau$ value by a small amount. This was executed 
separately for each of the relevant physical parameters, like temperature and velocity, 
by adding 1\,K and 10\,m\,s$^{-1}$, respectively, to the initial model. 
The results differ from model to model. In our case, we used the HSRA model as the 
reference atmosphere. As described before, we only selected the inner wings and
line core of the \ion{Ca}{II} line ($-$0.8\,$\leq$\,$\Delta \lambda$\,$\leq$\,1.2\,\AA\ 
from the line core). 
For the LOS velocity, 
high values of the RF were detected 
between $-$4.2\,$\geq$\,$\log\tau$\,$\geq$\,$-$5.8, corresponding to the chromosphere.
In addition, an enhanced RF was retrieved in the upper photosphere within the height range 
$-$2.7\,$\geq$\,$\log\tau$\,$\geq$\,$-$3.9. 
In the RFs for the temperature, high values were encountered 
in the chromosphere between $-$4.4\,$\geq$\,$\log\tau$\,$\geq$\,$-$5.2.  
Moreover, the spectral line is sensitive to temperature changes in the photosphere between 
$-$1.0\,$\geq$\,$\log\tau$\,$\geq$\,$-$3.7
(with a gradually decreasing RF with height).
In line with \citet{quinteronoda16}, we found a lower response of the \ion{Ca}{ii} line at 
around $\log \tau=$\,$-4$ in both RFs.

%--------------------------------------------------------------------
\subsection{Line-of-sight velocities}\label{Sect:LOSvelLorentzian}
In addition to the height-dependent LOS velocities inferred with NICOLE, we also performed Lorentzian fits 
with four coefficients to the \ion{Ca}{ii} 8542\,\AA\ line core between $-$0.8\,$\leq$\,$\Delta \lambda$\,$\leq$\,0.8\,\AA\
from the line center following the method described by \citet{gonzalezmanrique16}.
The four coefficients represent the peak value, centroid, half-width at half-maximum, and
a constant related to the continuum.
This technique is not as reliable as carrying out 
spectral line inversions \citep{socas06} since the shape of the \ion{Ca}{ii} 8542\,\AA\ line undergoes 
significant changes in active regions. However, a comparison between this fast technique and the time-consuming
inversions provides feedback on the reliability of this method.

\begin{figure}[!t]
 \centering
 \includegraphics[width=\hsize]{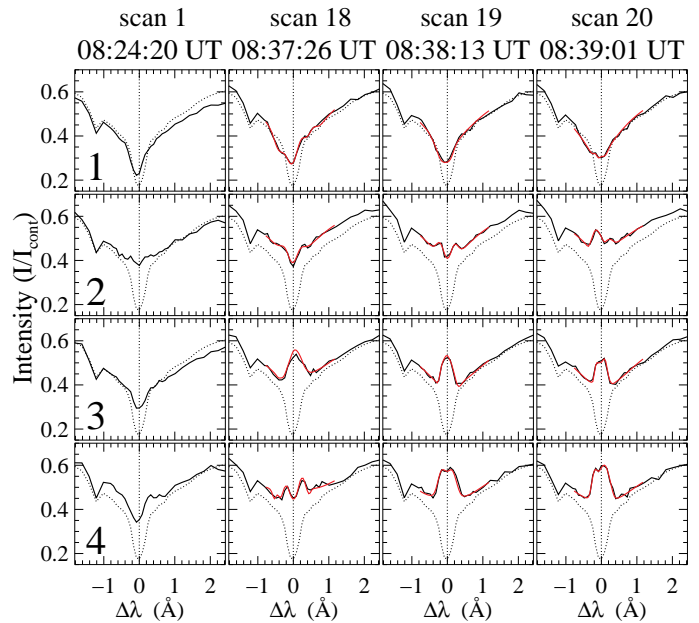}
\caption{Evolution (\emph{left} to \emph{right}) of the \ion{Ca}{ii} 8542\,\AA\ intensity 
          profile (solid black line) displayed for the most conspicuous scans (1 and 18 to 20) in four selected areas
          (1--4, \emph{top} to \emph{bottom}). 
          The four areas are also marked in all other figures. 
          Overplotted in solid red is the best fit from the NICOLE inversions in the spectral range 
          $-$0.8\,$\leq$\,$\Delta \lambda$\,$\leq$\,1.2\,\AA. The same pixel was chosen for each scan. 
          The time corresponds to the midpoint of each scan.
          For reference, the dotted profile shows the average quiet-Sun profile. }
  \label{Fig:CaIIprofiles}
\end{figure}

% =====================================================================
\section{Results}
% =====================================================================
We concentrated our investigations on small brightening events detected at 08:37\,UT as an enhancement
of the \ion{Ca}{ii} 8542\,\AA\ line-core intensity in the GFPI data.
The brightenings are marked with numbers 2--4 in all figures.
We closely follow
their origin and evolution in all AIA wavelength bands as well as with HMI LOS magnetograms. 
The \ion{Ca}{ii} spectra provide valuable information about the temperatures
and LOS velocities inferred with the NICOLE inversion code.

%--------------------------------------------------------------------
\subsection{Ca\,\textsc{ii} 8542\,\AA\ line-core enhancements}
The observations targeted a $\sim$\,10\arcsec\ wide pore with a rudimentary penumbra protruding from the 
left-hand side of the central part of the pore. 
A continuum image is shown in the bottom panel of Fig. \ref{Fig:overview}. 
The contours outline the positive (red)
and negative (blue) polarities taken from the HMI magnetogram clipped at $\pm$\,350\,G (shown in the 
bottom right panel of Fig.~\ref{Fig:overview} ). 

During the about 16-minute observations, the penumbra was 
slowly evolving, but no expansion was evident. 
However, in the third to last scan (scan 18), we observed a sudden spectral enhancement of the \ion{Ca}{ii} line-core 
intensity at two spatial locations with sizes of about 1\arcsec\,$\times$\,1\arcsec and 2\arcsec\,$\times$\,2\arcsec, 
areas 3 and 4 in the GFPI \ion{Ca}{ii} panel in Fig. \ref{Fig:overview}. 
The scan started at 08:37:03\,UT, and
the line core of \ion{Ca}{ii} (which was sampled a few seconds after the start of the scan) appeared in emission.
Approximately 1.5 minutes later, in the last scan (scan 20), a third  
location (area 2) clearly showed \ion{Ca}{ii}
line-core emission. The three brightening areas are overplotted as white or green
contours in Fig.~\ref{Fig:overview}. 
In addition, area 1 stands for a darker filamentary
structure, with deep \ion{Ca}{ii} absorption, seen in the bottom GFPI panel, which we discuss below. 
The area marked with a triangle in the GFPI panels 
(upper left corner, at coordinates $(x, y)\sim$(8\arcsec,31\arcsec))
also presents an enhancement of the 
\ion{Ca}{ii} line core. However, in contrast to the brightenings in areas 2--4, the enhancement is persistent 
during the 20 scans. 

A closer inspection of the spectral profiles is plotted in Fig.~\ref{Fig:CaIIprofiles}. 
Each row shows one representative profile from areas 1--4.  
For reference, the quiet-Sun spectrum (dotted line) is also shown. 
It results from computing the median of 50\,000 profiles in a quiet-Sun 
area on the right-hand side of the FOV. 
The first column corresponds to the first scan with the GFPI (08:24\,UT). It is the farthest away in time
($\sim$13\,min) from the 
beginning of the brightenings and hence represents a quiet atmosphere. 
The next three columns show the start of the brightenings.  
In the following, we call this nascent emission of the line core the ``impulsive phase''. 
While areas 3 and 4 reach the impulsive phase already in scan 18, area 2 reaches a clear line-core emission 
later, as is best seen in scan 20.
The \ion{Ca}{ii} impulsive phase lasts at least 1.5 minutes. Unfortunately, no additional 
observations were taken after scan 20. 

In area 1, that is, in the dark filamentary structure, the intensity profiles appear blueshifted in all 
scans (top panels of Fig.~\ref{Fig:CaIIprofiles}). The spectrum becomes significantly
asymmetric in scan 18, which corresponds to the beginning of the impulsive phase.
Importantly, the \ion{Ca}{ii} line core does not turn into a clear emission 
at any time in this area during the temporal period we observed. 
For a further understanding of this impulsive event, we discuss these findings in the
context of space data from SDO in the next sections.

\begin{figure}[!t]
 \centering
 \includegraphics[width=\hsize]{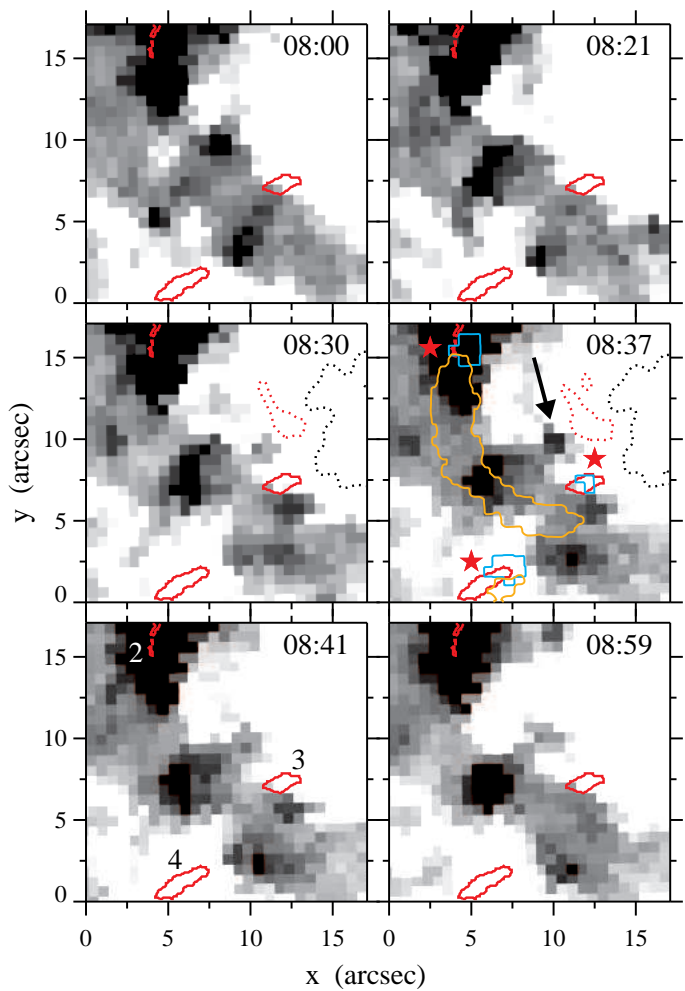}
 \caption{HMI LOS magnetograms clipped between $\pm$\,40\,G at selected times from 08:00 until 09:00\,UT.
 The FOV corresponds to the black rectangle in the magnetogram panel of Fig. \ref{Fig:overview}. The 
  contours show the following: blue represents AIA 1600\,\AA\ brightenings, 
  orange marks the flaring arch in AIA 131\,\AA,\, 
  red shows the GFPI \ion{Ca}{ii} 8542\,\AA\ line-core brightenings, and dotted black and red represents  
 the pore and forming penumbra, respectively. All contours refer to their time
 except the \ion{Ca}{ii} brightenings, which are static in all panels.
 The arrow at 08:37\,UT points toward a newly appearing negative polarity. The three red stars   
 next to the red contours
 indicate the beginning and location of the impulsive phase.}
  \label{Fig:magnetograms}
\end{figure}

%--------------------------------------------------------------------
\subsection{AIA response to the impulsive event}
We visually inspected all AIA wavelength bands and 
focused on the time between 
08:35--08:42\,UT (some channels are available as an online movie 
between 08:35:08 and 08:39:08\,UT). In addition, 
we computed light curves of a smaller region of interest (not shown). 
A sudden and strong single brightening patch 
of about 2\arcsec\,$\times$\,2\arcsec\ was clearly detected in all AIA images at 08:37\,UT 
(the exact position is marked with an arrow with letter `a' in the AIA 1600\,\AA\ panel of Fig.~\ref{Fig:overview}). 
The position partially overlaps with brightening area 4 from the GFPI.
Its lifetime was very short (<\,1\,min). 
Similar to the \ion{Ca}{ii} line-core enhancements of the
GFPI, 
the AIA 1600\,\AA\ images also show three areas (almost cospatial with the \ion{Ca}{ii} areas) 
that suddenly became brighter simultaneously
(three arrows in Fig.~\ref{Fig:overview}), but in contrast to \ion{Ca}{ii} 
(>\,1.5\,min), quickly disappeared. 
Only a small increase in intensity was found in AIA 1700\,\AA\ images 
when thoroughly inspecting the light curve. 
In contrast, all other wavelength 
bands exhibited steep increases of their intensity. 
In the time range of 12--24\,s after the strong single brightening patch, 
a double arch-shaped brightening 
became visible in all AIA channels except at 1600 and 1700\,\AA\ (Fig.~\ref{Fig:overview}).
These two wavelengths belong substantially to lower heights in the solar atmosphere 
\citep[Table 1 in][]{aia}. Except for a slight shift, the very bright arch is almost cospatial to the
dark filamentary structure of area 1 in the \ion{Ca}{ii} line-core images. 
According to the AIA channels that sample the hottest layers, 
the flaring arches were visible for a duration of about 3.0--3.5\,minutes. Their length
was around 20\arcsec. Spontaneous smaller post-flaring events were subsequently detected.
Interestingly, AIA 304\,\AA\ images already showed occasional bright areas 
at least 30 minutes before the impulsive event of 08:37\,UT. The other 
coronal wavelength bands also showed occasional but fainter brightenings outside of the described region of interest.

\begin{figure*}[!t]
\resizebox{\hsize}{!}{\includegraphics{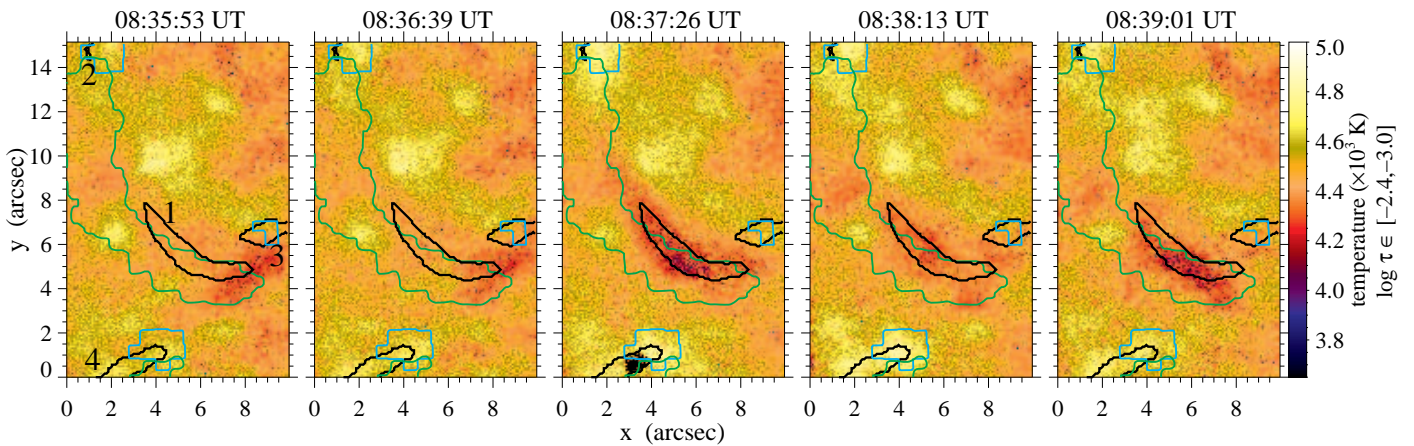}}
\caption{Average temperature scans 16--20 
         (\emph{left} to \emph{right}) inferred from the NICOLE inversions 
         of the \ion{Ca}{ii} 8542\,\AA\ line between the heights of $\log \tau \in [-2.4,-3.0]$. 
         The time corresponds to the midpoint of each scan sequence. The impulsive phase starts at 08:37\,UT.
         Overplotted black contours, marked with numbers 2--3 in the left-hand panel, show cumulative GFPI \ion{Ca}{ii} line-core enhancements between 08:37--08:39\,UT. The black contour marked with 1 encompasses 
         a dark filamentary structure seen in the GFPI \ion{Ca}{ii} line-core images. 
         AIA 131 (green) and 
         AIA 1600\,\AA\ (blue) contours show brightenings at 08:37:56 and 08:37:28\,UT, respectively.
         The black area at the bottom of the panel at 08:37:26\,UT does not show information
         on the inversions owing to bad fits ($\chi^2$ exceeds threshold criterion).
         The FOV corresponds to the black rectangles shown in the GFPI panels of Fig. \ref{Fig:overview}.}
        \label{Fig:temperatures}
\end{figure*}

%--------------------------------------------------------------------
\subsection{HMI magnetograms}
Selected HMI LOS magnetograms between 08:00--09:00\,UT
are depicted in Fig. \ref{Fig:magnetograms}. 
A smaller region of interest was chosen (rectangle in the magnetogram panel of Fig. \ref{Fig:overview}).
The magnetograms were clipped at $\pm 40$\,G 
to better visualize very faint photospheric magnetic-field concentrations. 
A concentration of
negative (black) polarity surrounded by positive (white) polarities was present on all sides.
The red solid contours are permanently present in all panels and encompass \ion{Ca}{ii} profiles
with line-core enhancements between 08:37--08:39\,UT (same as the white or green contours in Fig. \ref{Fig:overview}). 
The other contours vary according to their time stamp.  
Dynamic changes in central negative polarity were seen between 08:00 and 08:21\,UT in the top panels
of Fig. \ref{Fig:magnetograms}. 
A few minutes later, at 08:30\,UT, a negative-polarity 
patch gradually grows next to the right-hand solid red contour (area 3). The panel at 08:37\,UT is highlighted
by three red stars next to the contours indicating where the impulsive event started according to the
\ion{Ca}{ii} spectra. 
The three brightening areas in the AIA 1600\,\AA\ 
channel are outlined with blue contours. They match the \ion{Ca}{ii} areas well.
Immediately after this brightening, the most prominent 
flaring arch appeared in the hotter AIA channels, which was outlined by the orange contour 
in the AIA 131\,\AA\ filtergram. Another flaring arch (not shown) connected the upper and lower left-hand
red stars. This is more distinct in images of AIA 94\,\AA\ (see Fig. \ref{Fig:overview}), but also in all
AIA wavelength bands except AIA 1600 and 1700\,\AA. Furthermore, the 08:37\,UT panel is overlaid with 
three dotted contours: the black contour outlines the pore and the red contours the forming penumbra. They were 
determined from the GFPI broad-band images. 
Interestingly, next to the forming penumbra, a patch of negative polarity invaded
the extensive positive polarity (see black arrow in Fig. \ref{Fig:magnetograms}). This intrusion became visible 
at 08:30\,UT, was most prominent at 08:37\,UT, lasted until 08:39\,UT, and then slowly faded away. 
Although the negative polarity almost disappeared, the 
positive polarity did not fill the gap. 
However, a direct relation of this intrusion or the forming penumbra to 
the appearance of the brightenings cannot be established. Nevertheless, 
it indicates that the photospheric magnetic field was undergoing changes during this phase. 

After overplotting the three areas of \ion{Ca}{ii} line-core enhancements (areas 2--4) onto the magnetograms,
it becomes evident that each one shows different magnetic properties. One is rooted in the positive polarity
(area 4), the other in the negative polarity (area 2), and the third touches mixed polarities, 
although mainly positive (area 3). 
The double arches seen in all high-energy AIA channels connected the negative polarity  
(area 2) with the positive (area 4) and mixed (area 3) polarities.

\begin{table*}[!ht]
\caption{Evolution of average temperature $\langle T \rangle$ and standard deviation $\sigma_T$ inside of areas 
1--4 that are 
marked in Fig. \ref{Fig:temperatures}.  }    
\label{tab:temp}      % is used to refer this table in the text
\centering                          % used for centering table
    \begin{tabular}{c | c | c c c c c | c c c c c}        % centered columns (4 columns)
\hline\hline                 % inserts double horizontal lines
%Area                  &      & \multicolumn{5}{c}{$\log \tau \in [-2.4,-3.0]$} & \#   \\    
    &  &\multicolumn{5}{c}{$\log \tau \in [-4.6,-5.2]$ ~($z_2$)}& \multicolumn{5}{c}{$\log \tau \in [-2.4,-3.0]$ ~($z_1$)} \rule[-6pt]{0pt}{18pt}\\
\hline      
    & Area &                 &      & Scan &      &           & \multicolumn{5}{c}{Scan} \rule{0pt}{11pt}    \\  
    &      &             16  &  17  &  18  &  19  &  20       &              16  &  17  &  18  &  19  &  20 \rule[-4pt]{0pt}{10pt}\\
\hline
%                  & \multicolumn{10}{c}{Area 1}             \\
%\hline                        % inserts single horizontal line
 $\langle T \rangle$ (K)     &   & 5156  & 5199  & 5278  & 5194  & 5297  & 4408 & 4418 & 4283 & 4353 & 4313\rule{0pt}{11pt}   \\  
 $\sigma_T$ (K)& 1 &  \pha 801  &  \pha 811  &  \pha 837  &  \pha 899  & 1056  &   \phb 85 &   \phb 92 &  \pha 117 &   \phb 85 &  \pha 120  \\
 \#            & \rule[-4pt]{0pt}{10pt}  &  \pha 982  &  \pha 975  &  \pha 933  &  \pha 963  &  \pha 958  &  \pha 982 &  \pha 975 &  \pha 933 &  \pha 963 &  \pha 958  \\
 \hline
%               &   & \multicolumn{10}{c}{Area 2}              \\
%\hline                        % inserts single horizontal line
 $\langle T \rangle$      &    & 5627  & 5383  & 5777  & 5844  & 6436  & 4566 & 4593 & 4669 & 4619 & 4632 \rule{0pt}{11pt}    \\  
 $\sigma_T$ & 2  & 1195  &  \pha 973  & 1445  & 1546  & 1572  &  \phb  69 &  \phb  56 &  \phb  55 &  \phb  79 &  \phb  73     \\
 \#         &  \rule[-4pt]{0pt}{10pt}  &  \phb  63  &   \phb 59  &  \phb  63  &  \phb  59  &  \phb  63  &  \phb  63 & \phb   59 &  \phb  63 &  \phb  59 &  \phb  63  \\
\hline
%            &      & \multicolumn{10}{c}{Area 3}              \\
%\hline                        % inserts single horizontal line
 $\langle T \rangle$      &   & 5569  & 5598  & 6218  & 6183  & 6447  & 4463 & 4453 & 4533 & 4460 & 4472\rule{0pt}{11pt}    \\  
 $\sigma_T$ & 3  & 1220  & 1550  & 1611  & 1717  & 1654  &   \phb 88 &  \pha 101 &   \phb 86 &   \phb 94 &   \phb 77   \\
 \#         & \rule[-4pt]{0pt}{10pt}     &  \pha 341  &  \pha 336  &  \pha 331  & \pha  320  &  \pha 336  & \pha  341 &  \pha 336 &  \pha 331 &  \pha 320 &  \pha 336    \\
\hline
%            &       & \multicolumn{10}{c}{Area 4}             \\
%\hline                        % inserts single horizontal line
 $\langle T \rangle$       &   & 5590  & 5591  & 5900  & 6402  & 6498  & 4604 & 4627 & 4705 & 4714 & 4658 \rule{0pt}{11pt}    \\  
 $\sigma_T$  & 4 & 1408  & 1695  & 1694  & 1773  & 1918  &  \phb 87 &   \phb 86 &  \pha 104 &  \pha 103 &   \phb 98     \\
 \#          & \rule[-4pt]{0pt}{10pt}    &  \pha 434  &  \pha 426  & \pha  329  &  \pha 402  & \pha  415  &  \pha 434 &  \pha 426 &  \pha 329 &  \pha 402 &  \pha 415     \\
\hline             
\end{tabular}
\tablefoot{
The row with the hashtag shows the number of pixels inside of each area. Note that the number varies 
because failed inversions were excluded (see threshold criterion in Sect. \ref{Sect:inversions}).
The \emph{left} and \emph{right} blocks correspond to two different heights ($z_1$ and $z_2$), 
$\log \tau \in [-4.6,-5.2]$ and $\log \tau \in [-2.4,-3.0]$, respectively.
}
\end{table*}

\begin{figure*}[!t]
   \resizebox{\hsize}{!}{\includegraphics{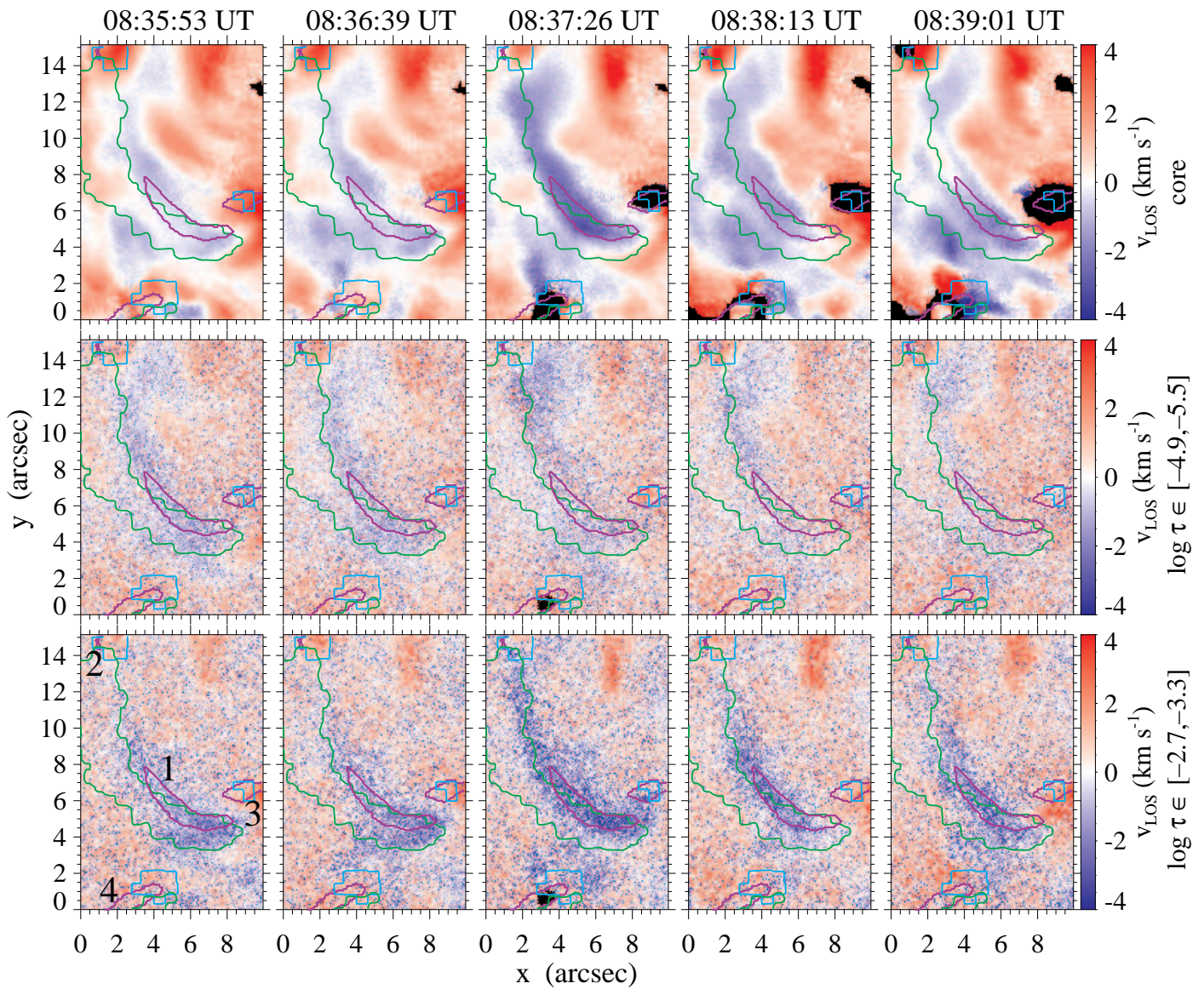}}
      \caption{\emph{Top} row: LOS velocities retrieved from line-core Lorentzian fits. 
      The \emph{middle} and \emph{bottom} panels show averaged velocities inferred with 
      NICOLE between $\log \tau \in [-4.9,-5.5]$ and $\log \tau \in [-2.7,-3.3]$, respectively. 
      The contours are the same as in Fig. \ref{Fig:temperatures}, but switching black by purple.
      The impulsive phase starts at 08:37\,UT.}
         \label{Fig:velocities}
\end{figure*}

%--------------------------------------------------------------------
\subsection{Temperature}
We derived temperatures at different heights from the inversions of the \ion{Ca}{ii} 8542\,\AA\ spectra. 
We concentrated on a small region 
(black rectangle in the GFPI panels of Fig. \ref{Fig:overview}) that included the areas  
of interest (1--4) in the last five scans of the GFPI. In this way, we covered the pre-impulsive and 
impulsive phases. 
Atmospheric fluctuations on scales shorter than the photons mean-free path are irrelevant for the radiative transfer.
Since we are not interested in such small-scale spurious fluctuations, the atmospheric parameters were averaged
over a range of $\log \tau = 0.7$. In particular, Figure~\ref{Fig:temperatures}
shows temperatures averaged over a range of $\log \tau \in [-2.4,-3.0]$, 
which corresponds to the upper photosphere. All panels are scaled between 
3650 and 5000\,K. 
The most relevant events were overplotted as contours:
the \ion{Ca}{ii} line-core emission in areas 2--4 between 08:37--08:39\,UT (black), 
the AIA 1600\,\AA\ brightening at 08:37\,UT (blue), 
and the flaring arch seen in AIA 131\,\AA\ at 08:37\,UT (green).
Since area 1 had no line-core emission profiles, we set up a different criterion. 
We chose the most blueshifted profiles of the \ion{Ca}{ii} line by
using filtergrams at $\Delta \lambda = -0.32$\,\AA\ from line center 
at 08:37\,UT. For the contour, we selected profiles with the lowest intensity, 
that is, the deepest line core. 
Although the flaring arch contour from AIA 131\,\AA\
largely covers area 1, there is a small offset with respect to relevant temperature (and
as seen later in the case of LOS velocity) changes. 

The first two panels of Fig. \ref{Fig:temperatures} show the temperatures before the impulsive phase. 
The middle panel (08:37\,UT) represents the beginning of the impulsive event. As expected, 
the temperature increases in areas 2--4. 
Conversely, the temperature in area 1 decreases. 
In the last panel, the temperature decreases gradually in areas 2--4, although the temperature decrease is fastest in area 3. 
The central area 1 remains cooler than in the pre-impulsive phase. 
Interestingly, area 1 shows an opposite behavior to the coronal
channels of AIA, where brightenings appear in the same location (green contour),
indicating increasing temperatures. 

From a quantitative point of view, we summarized the temperature variations of the
four areas in Table \ref{tab:temp}. First, we concentrate on the second half of the columns, which belong
to average temperatures between $\log \tau \in [-2.4,-3.0]$ (shown in Fig. \ref{Fig:temperatures}).
The table exhibits the average and standard deviation of the temperature and the amount of pixels
inside of each area. Scans 16 and 17 were acquired before the impulsive event. The impulsive event 
starts with scan 18 and shows an increase of 70--103\,K in areas 2--4 with respect to scan 16. 
Area 4 seems to be even hotter in scan 19. Nevertheless, we cannot rule out that scan 18 might have a higher
temperature, but after ignoring pixels owing to bad fitting, the computed average temperature is lower. 
Area 1 behaves in a different way. At the beginning of the impulsive phase, the 
average temperature drops 
by 125\,K with respect to scan 16. In the last two scans, the temperature increases and then decreases
again, respectively. 

We move farther up in the atmosphere and concentrate on a chromospheric height that shows
a high sensitivity to temperature perturbations in the \ion{Ca}{ii} 8542\,\AA\ line. This is shown
in the first half of the columns of Table \ref{tab:temp}, where we analyze the average temperature
variations within $\log \tau \in [-4.6,-5.2]$. We first note that the standard
deviation is one order of magnitude larger than in the  
$\log \tau \in [-2.4,-3.0]$ layer. In addition, the temperature itself is higher. In areas 2--4, 
the temperature increases by about 800--900\,K between scans 16 and 20. Furthermore, 
in contrast to what we reported for the temperature between $\log \tau \in [-2.4,-3.0]$, 
the temperature continues to rise after the impulsive phase, that is, after scan 18. 
The highest temperatures
occur in the last scan. Interestingly, at this height, area 1 shows a rise in temperature 
by about 140\,K between the first and last scans. Although this increase is much weaker than the increase 
reported for areas 2--4, it is consistent with the increasing temperature seen in the higher AIA channels
(see Fig.\ref{Fig:overview}).

%--------------------------------------------------------------------
\subsection{Line-of-sight velocity}\label{Sect:velocities}
Figure \ref{Fig:velocities} compares the LOS velocities inferred from the \ion{Ca}{ii} 8542\,\AA\ line
at three different atmospheric heights and along the five last GFPI scans. 
From top to bottom we are mapping from higher to lower layers of the atmosphere: 
line core (see Sect. \ref{Sect:LOSvelLorentzian}), $\log \tau \in [-4.9,-5.5]$, and $\log \tau \in [-2.7,-3.3]$.
The latter two are height ranges that are most sensitive to velocity variations according to the analyzed response
functions (see Sect. \ref{Sect:responsefunct}). The FOV and the contours are the same as 
shown for the temperature in Fig. \ref{Fig:CaIIprofiles}. The LOS velocities are scaled 
between $\pm$\,4\,km\,s$^{-1}$. Negative (positive) values denote blueshifts (redshifts)
of the \ion{Ca}{ii} line.

The top panels plot
the velocities retrieved from the Lorentzian fits of the line core.
They serve as a first approximation since they often fail when more complex
profiles appear. This is the case in the black areas in Fig. \ref{Fig:velocities}, 
which are mainly dominated by \ion{Ca}{ii} 
line-core emission profiles. In these regions, an inversion code such as NICOLE copes well with
the more complex profiles,
with the exception of few pixels in area 4 of the middle panel. After inspection, the line-core
velocities represent a combination of the respective velocity maps shown in the middle and bottom rows.
The velocities belonging to $\log \tau \in [-2.7,-3.3]$ (bottom row) show predominant blueshifts
in the middle and lower half of the FOV. Conversely, the $\log \tau \in [-4.9,-5.5]$ velocities
(middle row) show an enhancement of the blueshifts in the upper half of the FOV. Hence, the two heights extracted
from the NICOLE inversions complement each other.

\begin{table*}[!ht]
\caption{Same as Table \ref{tab:temp}, but for the average LOS velocities at
$\log \tau \in [-4.9,-5.5]$ and $\log \tau \in [-2.7,-3.3]$.}    
\label{tab:vel}      % is used to refer this table in the text
\centering                          % used for centering table
    \begin{tabular}{c | c | c c c c c | c c c c c}        % centered columns (4 columns)
\hline\hline                 % inserts double horizontal lines
%Area                  &      & \multicolumn{5}{c}{$\log \tau \in [-2.4,-3.0]$} & \#   \\    
     & &\multicolumn{5}{c}{$\log \tau \in [-4.9,-5.5]$ ~($z_2$)}& \multicolumn{5}{c}{$\log \tau \in [-2.7,-3.3]$ ~($z_1$)} \rule[-6pt]{0pt}{18pt}\\
\hline      
     & Area &             &      & Scan &      &           & \multicolumn{5}{c}{Scan}   \rule{0pt}{11pt} \\  
     & &              16  &  17  &  18  &  19  &  20       &              16  &  17  &  18  &  19  &  20 \rule[-4pt]{0pt}{10pt}\\
\hline
%                  & \multicolumn{10}{c}{Area 1}             \\
%\hline                        % inserts single horizontal line
 $\langle v_\mathrm{LOS} \rangle$ (km\,s$^{-1}$) & & $-$0.73 & $-$0.61 & $-$0.45 & $-$0.07 & $-$0.08 & $-$1.22 & $-$1.32 & $-$2.18 & $-$1.31 & $-$1.28\rule{0pt}{11pt}  \\
 $\sigma_v$ (km\,s$^{-1}$)      & 1 &    \phd 1.13 & \phd 1.26    &    \phd 1.42 &    \phd 1.44 &    \phd 1.43 &    \phd 1.15 &   \phd  1.30 &   \phd  1.72 &   \phd  1.49 &    \phd 1.63 \\
 \#                             & \rule[-4pt]{0pt}{10pt}  &     \phc 982 &  \phc 975    &  \phc 933 &     \phc 963 &    \phc  958 &     \phc 982 &    \phc  975 &  \phc 933 &    \phc  963 &     \phc 958 \\
 \hline
%                  & \multicolumn{10}{c}{Area 2}              \\
%\hline                        % inserts single horizontal line
  $\langle v_\mathrm{LOS} \rangle$ &  & \phd 0.38 &\phd  0.25 &\phd  0.38 &\phd  0.18 & \phd 0.64 & \phd 0.73 & \phd 0.72 & $-$0.22 & \phd 0.97 & \phd 0.32 \rule{0pt}{11pt} \\
 $\sigma_v$          & 2& \phd 1.46 &\phd  1.46 & \phd 1.27 & \phd 1.39 & \phd 1.45 & \phd 1.46 &\phd  1.25 &   \phd  1.18 & \phd 1.05 & \phd 1.68  \\
 \#                  &  \rule[-4pt]{0pt}{10pt} & \phe  63 &   \phe 59 &   \phe 63 &   \phe 59 &  \phe  63 &   \phe 63 &  \phe  59 &     \phe  63 &  \phe  59 &   \phe 63 \\
\hline
%                  & \multicolumn{10}{c}{Area 3}              \\
%\hline                        % inserts single horizontal line
  $\langle v_\mathrm{LOS} \rangle$ &  & \phd 0.04 & \phd 0.35 &\phd  0.75 & \phd 0.23 & \phd 0.25  & \phd 1.08 & \phd 0.91 & $-$0.75 & \phd 0.99 & \phd 1.12\rule{0pt}{11pt}  \\
 $\sigma_v$          & 3& \phd 1.81 & \phd 1.83 & \phd 1.76 & \phd 1.87 & \phd 1.69  & \phd 1.34 & \phd 1.69 &    \phd 1.92 & \phd 1.56 & \phd 1.72  \\
 \#                  & \rule[-4pt]{0pt}{10pt} & \phc  341 &  \phc 336 & \phc  331 &  \phc 320 &  \phc 336  & \phc  341 &  \phc 336 &     \phc 331 & \phc  320 &  \phc 336 \\
\hline
%                   & \multicolumn{10}{c}{Area 4}             \\
%\hline                        % inserts single horizontal line
 $\langle v_\mathrm{LOS} \rangle$ &  & \phd 0.72 &\phd \phd  0.67 &\phd  0.68 & \phd 0.75 & \phd 0.79 & \phd 0.28 & \phd 0.73 & $-$0.29 &\phd  0.33 & \phd 0.29\rule{0pt}{11pt}  \\
 $\sigma_v$         &  4& \phd 1.65 & \phd 1.66 & \phd 1.65 & \phd 1.56 & \phd 1.69 & \phd 1.71 & \phd 1.83 &    \phd 2.43 & \phd 2.03 &\phd  2.23  \\
 \#                 & \rule[-4pt]{0pt}{10pt} &  \phc 434 &  \phc 426 &  \phc 329 &  \phc 402 &  \phc 415 &  \phc 434 &  \phc 426 &    \phc  329 & \phc 402 &  \phc 415 \\
\hline             
\end{tabular}
\end{table*}

Figure \ref{Fig:velocities} 
gives indications about the dynamics within the region of interest. The contours of
area 1 (by definition represented by blueshifted \ion{Ca}{ii} profiles) and the flaring arch 
(green contour) show blueshifts during the five scans. In particular, the strongest blueshifts coincide with the
middle scan at 08:37. This is consistent with the impulsive phase and the flaring arch seen in 
the AIA wavelength bands. Moreover, a large part of the \ion{Ca}{ii} blueshifts match the
flaring arch. In contrast, areas 2--4 appear mainly redshifted. 
To support this qualitative finding, we present the average LOS velocities in Table \ref{tab:vel}.
The table follows the same layout as Table \ref{tab:temp}. We first concentrate
on the lowest height, which corresponds to $\log \tau \in [-2.7,-3.3]$. Area 1 consistently shows 
blueshifts of around $-$1.3\,km\,s$^{-1}$ before and after the impulsive event at 08:37\,UT. The average
velocity increases to $-$2.18\,km\,s$^{-1}$ during the impulsive phase. The remaining areas 2--4,
which are distributed around area 1, exhibit redshifts before and after the impulsive phase. 
During the eruption, the velocities become negative, thus representing blueshifts. 
However, these blueshifts between $-0.22$ and
$-0.75$\,km\,s$^{-1}$ are considerably lower than those seen in area 1.

Higher up in the chromosphere, between $\log \tau \in [-4.9,-5.5]$ 
(first half of the columns in Table \ref{tab:vel}), the plasma behaves
differently. The retrieved velocities in area 1
still show blueshifts, but with a gradual decrease between scans 16 and 20, from $-0.73$ 
to $-0.08$\,km\,s$^{-1}$, respectively. Furthermore, in the post-impulsive maps (after 08:37\,UT), 
this drop in velocity is most noticeable. 
This is also evident in the velocity maps in the middle row in Fig. \ref{tab:vel}. 
Areas 2--4 are completely dominated by redshifts. In particular, 
area 3 exhibits an increased downflow (redshift) during the impulsive phase (an increase of about 
0.70\,km\,s$^{-1}$ between scans 16 and 18). Area 4 manifests the most stable velocities, which
fluctuate around 0.73\,km\,s$^{-1}$.

An interesting observational feature occurs in the upper right part of the FOV in the 
$\log \tau \in [-2.7,-3.3]$ and line-core panels of Fig. \ref{Fig:velocities}. An elongated vertical
patch of redshifts is present at position (7\arcsec, 14\arcsec). A careful comparison with the GFPI panels
in Fig. \ref{Fig:overview} yields that this patch does not coincide with the filamentary structure
present in the FOV. 
The forming penumbra from the large pore starts where the red patch ends.
Therefore we speculate that flows might be interacting between these two structures. However, we do
not find any direct relation to our impulsive event.

% =====================================================================
\section{Discussion}
% =====================================================================

\subsection{Classification of the brightenings}
We presented a multiwavelength analysis motivated by sudden small brightenings
detected in \ion{Ca}{ii} 8542\,\AA\ line-core images. We identified three pronounced brightening areas (tagged with
numbers 2--4 in all figures). The brightenings belonged to a large active region and had extensions up to
2\arcsec\,$\times$\,2\arcsec. Next to the brightenings
was a large pore with a slowly forming penumbra. 
Two of the brightenings were rooted in two different polarities (areas 2 and 4),
while the third (area 3) resided 
next to mixed polarities, although the positive polarity dominated, in the photosphere. 
Seen in \ion{Ca}{ii} spectra, 
line-core emission first started in areas 3 and 4, indicated by a clear reversal 
of the line-core intensity (Fig.~\ref{Fig:CaIIprofiles}). Area 2 also showed an enhanced line center that
completely reversed in the last scan, about 1.5\,min after the reversal in areas 3 and 4. 
The line-core emission remained until the end of the observations. 
The exact duration of the brightening was unknown, however, but a lower limit of 1.5\,min can be established. 
At the same time, very close to these brightenings, three areas of about the same size flared up
simultaneously in the AIA 1600\,\AA\ channel (arrows in Fig.~\ref{Fig:overview} in the respective panel).
All other AIA channels showed only one single but very intense brightening patch in area 4
(faint patch for AIA 1700\,\AA), which 
belonged to a high concentration of positive polarity. A few seconds later, 
hot arch-shaped structures appeared simultaneously in all chromospheric and coronal AIA channels, 
producing sufficiently high temperatures to ionize and recombine
\ion{He}{ii} and Fe atoms (see online movie).  
Spontaneous smaller post-flaring events were also detected a few minutes later. 
The fact that the earlier brightenings in the AIA 1600\,\AA\ images lasted less than 24 seconds (time interval 
between AIA 1600\,\AA\ filtergrams) suggests that the brightenings belonged to the beginning of a microflare. 
We exclude EBs, IBs, and FAFs, since these events have substantially longer lifetimes in that wavelength band, even though \ion{Ca}{ii} 8542\,\AA\ line-core 
images show longer brightenings in time, at least 1.5 minutes, until the end of our observations. 

The observed microflare shares an interesting
feature with FAFs. Similar to the FAFs presented by \citet{vissers15}, our observations showed very bright 
flaring arches next to the brightenings that lasted between 3.0--3.5\,min (4\,min in their case). The arch lengths were also similar ($\sim$20\arcsec). Nevertheless, this might depend on the active region and 
the specific configuration of the magnetic field. Another unclear
aspect is whether FAFs present
spectral profiles with line-core emission. In \citet{vissers15}, FAFs seem to have an enhanced intensity profile 
in H$\alpha$, including the line core. The \ion{Ca}{ii} 8542\,\AA\ profiles shown in Fig.~\ref{Fig:CaIIprofiles} clearly  present line-core emission together with enhancements of the wings. The former fact makes them easily distinguishable from EBs, which only show strongly enhanced intensity line wings, but the core remains unaffected. 
We exclude IBs because they do not show brightenings in coronal
wavelengths \citep{peter14}. This is another reason why we rule out EBs.  

Further information on the nature of the studied brightenings might be provided by data from 
the Geostationary Operational Environmental Satellite system (GOES) or the 
Reuven Ramaty High Energy Solar Spectroscopic Imager \citep[RHESSI,][]{rhessi}. No 
increase in count rate was found in the GOES X-ray flux during our time range. 
However, the 6--12\,keV band
of RHESSI shows a sharp peak at exactly 08:37\,UT, cotemporal to the beginning of the impulsive 
phase in our data. The 6--12\,keV band is often used for flare detection algorithms \citep{christe08}
and is dominated by thermal emission \citep{benz02}. Thus, this
provides another argument favoring the presence of an MF in our data.

\subsection{Origin of the microflare}
Following our argumentation, we consider our brightenings as part of an MF. We combine the 
available information to characterize the beginning of this MF. 
For the sake of clarity, we define two different layers ($z_1$ and $z_2$)
to facilitate the interpretation of the height-dependent \ion{Ca}{ii} 8542\,\AA\ inversions. 
The velocities $v_{z_1}$ and $v_{z_2}$ refer to the mean LOS velocities in the intervals $\log \tau \in [-2.7,-3.3]$ and
$\log \tau \in [-4.9,-5.5]$, respectively. Similarly, the temperatures $T_{z_1}$ and $T_{z_2}$ are defined
as the averages in the intervals $\log \tau \in [-2.4,-3.0]$ and $\log \tau \in [-4.6,-5.2]$, respectively. 
The optical depth decreases with height. Therefore we identify $z_1$ with the upper photosphere
and $z_2$ with the chromosphere. 

\subsubsection{Footpoints}
The MF starts with one brightening that appeared simultaneously in the lowest and highest 
atmospheric layers of AIA, in area 4 in Fig.~\ref{Fig:overview}.
This already distinguishes the analyzed MF from others that only exhibit emission in chromospheric layers and 
later present emission in the corona \citep{brosius09}.
At the same time, the \ion{Ca}{ii} 8542\,\AA\ line-core images from GFPI show 
brightenings in areas 3 and 4, and roughly 1.5\,min later in area 2. 
Interestingly, one particular AIA 1600\,\AA\ 
filtergram shows the three brightenings at the beginning of the impulsive phase at 08:37\,UT. 
This is clearly seen 
in Fig.~\ref{Fig:magnetograms} in the 08:37\,UT panel. In this figure, the red contours represent the 
\ion{Ca}{ii} line-core brightenings of areas 2--4, while the blue contours outline the AIA 1600\,\AA\ intensity enhancements.
The two contours match fairly well (without taking into account projection effects). Therefore we consider areas 2--4 as the footpoints of the MF in the upper photosphere. They are rooted in opposite polarities.
We conjecture that the mixed polarity patch at area 3, which emerged a few minutes
before the beginning of the impulsive phase (Fig.~\ref{Fig:magnetograms}), might have triggered magnetic field reconnection. Our observations favor a scenario with reconnection, 
which leads to sudden energy release. 
Whether this reconnection occurs in the photosphere or
in higher layers of the atmosphere remains unclear from the available HMI magnetograms.
Different polarities belong to the three brightenings, but the magnetograms before and after the beginning of the MF 
do not show a clear pattern of magnetic reconnection in the photosphere. 

We now discuss the average LOS velocity changes associated with the footpoints inferred from the
\ion{Ca}{ii} 8542\,\AA\ inversions. Before the impulsive phase, downflows dominate in all three areas at both
heights (Fig.~\ref{Fig:velocities}). 
These downflows are small, typically below 1.1\,km\,s$^{-1}$ (Table \ref{tab:vel}). We see a fundamental
change in the upper photosphere at $z_1$ when the MF starts. All velocities in the three areas change their sign, 
switching from redshifts to
blueshifts. The strongest upflows of $v_{z_1}=-0.75$\,km\,s$^{-1}$ are seen in area 3. This is the
only area embedded in opposite magnetic polarities in the magnetograms. 
Conversely, higher in the atmosphere at $z_2$, the downflows persist. This is consistent with 
bidirectional flows like those reported by \citet{hong16}. However, their MF showed much higher 
LOS velocities, between $\pm 50$\,km\,s$^{-1}$, inferred also from the near-infrared \ion{Ca}{ii} line. 
The reason is probably that their MF was of GOES B-class, much more energetic than the MF
studied in this work.
The downflows at $z_2$ continue in the last two scans. 
After the impulsive phase, $v_{z_1}$ returns to redshifts in all areas. We conclude that 
at the footpoints the impulsive phase of the MF 
produces sudden upflows of plasma from $z_1$ into higher atmospheric layers. 
However, these upflows are not seen above at $z_2$. The origin of the bidirectional flows is unclear. 
A possible explanation is plasma evaporation and condensation. Nevertheless, inflows and outflows 
as a result of magnetic reconnection cannot be ruled out.

The temperature variations in the footpoints behave differently. All three areas show increases in their 
average temperature at both heights during the impulsive phase of the MF. 
Still, the temperature rise is much more noticeable at $z_2$, that is, up to $\sim$\,600\,K compared to $\sim$\,80\,K
at $z_1$. Area 3 showed the largest increment at both heights. 
Interestingly, $T_{z_2}$ continues to rise after the impulsive
phase in all three areas, while $T_{z_1}$ slowly decreases (Table~\ref{tab:temp}). 
The larger and steeper temperature gradients in $T_{z_2}$
compared to $T_{z_1}$ argues in favor of placing the origin of the MF in higher layers. 
There is no clear evidence from the \ion{Ca}{ii} 8542\,\AA\ intensity spectra that the MF originates
in the chromosphere. However, the consequences, such as temperature increments, are well detected there.

\subsubsection{Flaring arches}
The double arch-shaped brightenings seen in the AIA channels connect area 2 
in the upper part of the FOV with areas 3 and 4 (Fig.~\ref{Fig:overview}). 
Hence, the flaring arches connect the three footpoints. The 
\ion{Ca}{ii} 8542\,\AA\ line-core images show a dark filamentary structure (area 1) fairly cospatially with the 
flaring arches (without taking projection effects into account).  
This filamentary structure is dominated by slow upflows at $z_1$ and $z_2$ (see area 1 in Table \ref{tab:vel}). 
During the MF, the upflow becomes stronger, from $v_{z_1} = -1.32$ to $-2.18$\,km\,s$^{-1}$. This is not evident at $z_2$. 
The MF produces a lift-off of the plasma with moderate velocities at $z_1$, that is., 
in the upper photosphere. 
However, the filamentary structure does not disappear, it remains visible in the \ion{Ca}{ii} 8542\,\AA\ line-core images 
after the MF. 
We speculate that uplifted plasma from area~1 during the impulsive phase might drain 
down along the flaring arches toward the footpoints after the MF. 
This explains the increased downflows in some footpoints at both heights after the MF (Table~\ref{tab:vel}). 

While the plasma in area 1 moves up during the MF, the average temperature $T_{z_1}$ decreases by about 
130\,K. In contrast, $T_{z_2}$ increases $\sim$\,80\,K. An increment in $T_{z_2}$ is consistent with the 
temperature enhancements detected in the hotter AIA channels. 

Unfortunately, we are missing some part of the larger picture. This becomes clear as the cadences of AIA (12\,s and 24\,s) 
and the magnetograms of HMI (45\,s) are not fast enough to reproduce the rapid evolution of the phenomena
analyzed in this work. In addition, the long cadence of 47\,s for the \ion{Ca}{ii} 8542\,\AA\ observations needs to be
shortened in the future to capture rapid solar events. 
Higher time resolution is a great challenge that needs to be addressed by the new generation of solar telescopes, 
such as the Daniel K. Inouye Solar Telescope \citep[DKIST,][]{tritschler16} and the European Solar Telescope 
\citep[EST,][]{matthews16}.

\subsection{Line-of-sight velocity measurements}
Two different techniques were used to determine the LOS velocities from the \ion{Ca}{ii} 8542\,\AA\ line. 
On the one hand, we used Lorentzian fits with four coefficients to approximate the spectral profile. 
This method ensured a fast computation of the line-core velocities. However, the fits failed when complex
\ion{Ca}{ii} 8542\,\AA\ profiles were present (see black areas in Fig.~\ref{Fig:velocities}). 
Moreover, there is no height-dependent information. On the other hand, the spectral-line inversion
code NICOLE was used. It provided LOS velocities (and temperatures) at different heights in the atmosphere. 
Today, its main drawback is the computational power and time needed to invert the large number of profiles. 
Fortunately, the code has been parallelized recently, which makes the calculations substantially faster 
when high-performance computer clusters are available. 

A comparison between both methods is shown in Fig.~\ref{Fig:velocities}. 
Line-core fits to \ion{Ca}{ii} 8542\,\AA\ are not sufficient to study the physical quantities
of fast-evolving events on the Sun. They are useful as quick-look information, but often miss the details encoded in
complex spectral profiles. This is seen when comparing the two methods. The inversions reveal 
different velocities depending on the height. 
Especially at the footpoints, the velocities at $z_1$ and $z_2$ behave differently (Table~\ref{tab:vel}). 
The detection of bidirectional flows was extracted from the inversions and is not detectable in the Lorentzian fits. 
For the \ion{Ca}{ii} 8542\,\AA\ line, we therefore recommend using
more complex codes, such as inversion codes, to interprete the LOS velocities
in fast events.

% =====================================================================
\section{Summary and conclusions}
% =====================================================================
We carried out a multiwavelength analysis of 
small brightening events in the solar atmosphere based on high-resolution spectroscopic 
images of the GFPI at GREGOR in the \ion{Ca}{ii} 8542\,\AA\ line. 
We combined the ground-based observations with images from AIA and HMI.
The brightenings (areas 2--4 in all figures) were observed during their formation process and had sizes of
up to 2\arcsec\,$\times$\,2\arcsec. 
We excluded EBs as well as IBs as an explanation of these brightenings. Several reasons supported this 
rejection: 
(1) our brightenings are clearly seen with AIA across all layers of the atmosphere up to the corona, while
EBs and IBs do not have coronal counterparts, 
(2) our three brightenings were very short lived in AIA 1600\,\AA\ filtergrams
(<\,24\,s), while EBs have lifetimes of at least several minutes, and 
(3) the \ion{Ca}{ii} 8542\,\AA\ data showed emission in the
line core, whereas spectra of EBs only show largely enhanced line wings, but no line-core emission.
The scenario of a microflare appropriately fits our observations, although similarities to the FAFs
reported by \citet{vissers15} were found. Like in their investigation, bright flaring arches were detected 
next to our brightenings. Furthermore, the lifetimes of these flaring arches were about 3.0--3.5\,min with a
size of $\sim$20\arcsec. Simultaneous RHESSI observations showed
a sharp increase in count rate in the 6--12 keV band, which further confirmed our event as a MF.

The three brightenings 2--4 in the \ion{Ca}{ii} 8542\,\AA\ line-core images were considered as the footpoints
of the MF, whereas area 1 belonged to the flaring arches seen in AIA. 

Height-dependent inversions of the \ion{Ca}{ii} 8542\,\AA\ intensity spectra with NICOLE revealed LOS velocities 
and temperatures before and during the impulsive phase of the MF. We focused on two specific height ranges, 
the upper photosphere ($z_1$) and the chromosphere ($z_2$).

\begin{itemize}
 \item The MF heated the three footpoints at both heights. However, the temperature gradients during the impulsive phase 
 were much steeper at $z_2$ ($\sim$\,600\,K) than in $z_1$ ($\sim$\,80\,K). After the impulsive phase, 
 the temperatures at $z_2$ continued to rise at the footpoints, 
 whereas they decreased at $z_1$. 
 
 \item The LOS velocities at the footpoints at both heights were dominated by downflows. However, during the 
 impulsive phase, the velocities turned into blueshifts (up to $-0.75$\,km\,s$^{-1}$) in the upper photosphere at $z_1$. 
 At $z_2$ , the velocities remained redshifted during the MF. This indicates bidirectional flows along the LOS.
 
 \item As a consequence of the MF, flaring double arches appeared in all AIA channels between the footpoints,
 except in 1600 and 1700\,\AA\ filtergrams. One of the flaring arches spatially coincided with a dark 
 filamentary structure seen in the \ion{Ca}{ii} line-core image (area 1 in Fig.~\ref{Fig:overview}).
 From the inversions, we find that the plasma in the filamentary structure was predominantly moving upward
 before the MF occurred. During the impulsive phase, the plasma increased the upward motion at $z_1$ , reaching average LOS velocities
 of $-2.18$\,km\,s$^{-1}$. After the impulsive phase, the plasma motion returned to the initial 
 values for $z_1$, and at $z_2$ the velocities dropped almost to zero. This scenario is consistent with rising plasma 
 between the footpoints (along the filamentary structure seen in the \ion{Ca}{ii} line-core images), 
 during the impulsive phase. At the same time, plasma is draining downward along the footpoints.  
 
 \item The average temperatures in area 1 slightly increased ($\sim$\,80\,K) at $z_2$ during the impulsive phase, 
 which is consistent, but less than expected, with 
 the hotter AIA channels, which became brighter. Conversely,  the plasma became colder in the  
 upper photosphere at $z_1$ at the same time. 
 
\end{itemize}

We cannot answer the question where the microflare started.
For a rapid solar event like an MF, we would need a better temporal and spectral resolution to 
distinguish the time and height of the start of the MF. 
The \ion{Ca}{ii} 8542\,\AA\ diagnostics
suggest an origin above the \ion{Ca}{ii} line formation. The consequences of the MF are clearly reflected in the 
\ion{Ca}{ii} 8542\,\AA\ line, but they are less dramatic than
might be expected 
if magnetic reconnection occurred at this height.
Spectropolarimetric observations to infer the vector magnetic field in higher layers of the solar
atmosphere become crucial to answer this question.

% =====================================================================
\begin{acknowledgements}
The 1.5-meter GREGOR solar telescope was built by a German consortium under the leadership of the 
Kiepenheuer-Institut f\"ur Sonnenphysik in Freiburg (KIS), with the Leibniz-Institut f\"ur Astrophysik Potsdam (AIP), 
the Institut f\"ur Astrophysik G\"ottingen (IAG), the Max-Planck-Institut f\"ur Sonnensystemforschung in 
G\"ottingen (MPS), and the Instituto de Astrof\'isica de Canarias (IAC), and with contributions by the Astronomical 
Institute of the Academy of Sciences of the Czech Republic (ASCR).
The authors thank R. E. Louis for his help during the observations. 
Rob Rutten, Gregal J. M. Vissers, A. Warmuth, and H. Wang are greatly acknowledged for fruitful discussions. 
SJGM is grateful for financial support from the Leibniz Graduate School for Quantitative Spectroscopy in Astrophysics, a joint project of AIP and the Institute of Physics and Astronomy of the University of Potsdam, and he acknowledges support of the project VEGA 2/0004/16. Support from SOLARNET, a project supported by the European 
Commission's 7th Framework Program under grant agreement No. 312495, is greatly acknowledged.
Finally, we are grateful for helpful comments of the anonymous referee that improved the quality of this publication.
\end{acknowledgements}

%===============================================================================
%    BIBLIOGRAPHY
%===============================================================================

\bibliographystyle{aa}
\bibliography{aa-jour,biblio}

\end{document}